\documentclass[twocolumn,showpacs,preprintnumbers,amsmath,amssymb]{revtex4}

\begin{document}

\title{Disentangling positivity constraints for generalized parton
distributions}
\author{P.V. Pobylitsa}
\affiliation{
Institute for Theoretical Physics II, Ruhr University Bochum,
D-44780 Bochum, Germany\\
and Petersburg Nuclear Physics Institute, Gatchina, St. Petersburg, 188350,
Russia}

\begin{abstract}
Positivity constraints are derived for the
generalized parton distributions (GPDs) of
spin-1/2 hadrons. The analysis covers the full set of eight
twist-2 GPDs. Several new inequalities are obtained which constrain GPDs by
various combinations of usual (forward) unpolarized and polarized parton
distributions including the transversity distribution.
\end{abstract}

\maketitle

\section{Introduction}

It was shown during the last decade that a number of hard exclusive
processes,
including deeply virtual Compton scattering and hard exclusive meson
production,
are calculable in QCD in terms of generalized
parton distributions (GPDs)
also known as skewed, off-forward, or nondiagonal
\cite{MRGDH-94,Radyushkin-96,Ji-97,CFS-97,Radyushkin-97}.
GPDs cannot be measured directly: the cross sections of hard exclusive
processes give us information only about some integrals containing
combinations of GPDs with other perturbative and nonperturbative
functions.
Therefore any model independent theoretical constraints on GPDs are of interest.

The aim of this paper is to obtain new theoretical restrictions
on GPDs in terms of the usual (forward) parton distributions
using the so-called positivity bounds. The positivity
bounds for GPDs are derived by taking the square of the superposition
of parton-hadron states \cite{Martin-98,
Radyushkin-99, PST-99,Ji-98}:
\begin{equation}
\left\| \sum\limits_{k=1}^{2}\sum\limits_{\lambda \mu }c_{\lambda \mu
}^{(k)}\int \frac{d\tau }{2\pi }e^{i\tau x_{k}(P_{k}n)}\phi _{\mu }\left(
\tau n\right) |P_{k},\lambda \rangle \right\| ^{2}\geq 0\,\,.
\label{start-ineq}
\end{equation}
Here $|P_{k},\lambda \rangle $ is the nucleon state with momentum $P_{k}$ and
polarization $\lambda $, $\phi _{\mu }$ is the component of the quark (or
gluon) field corresponding to the polarization $\mu $, and $n$ is a
light-cone vector.

Expanding the LHS of inequality (\ref{start-ineq}) one obtains
an expression containing diagonal
terms corresponding to the usual parton distributions and nondiagonal
terms corresponding to GPDs.
Generally speaking, inequality (\ref{start-ineq}) should hold for any
coefficients $c_{\lambda \mu }^{(k)}$ but so far it has been analyzed only for
some specific sets of $c_{\lambda \mu }^{(k)}$. As a result
the positivity bounds for GPDs derived earlier are not as strong as
they could be.

In particular, an interesting inequality for GPD $H^q$ was derived in Refs.
\cite{Radyushkin-99, PST-99}
\begin{equation}
H^{q}(x,\xi ,t)\leq \sqrt{\frac{q(x_{1})q(x_{2})}{1-\xi ^{2}}}\,.
\label{Radyushkin-ineq}
\end{equation}
Here the notation of X. Ji \cite{Ji-98} is used for the quark GPDs
($H^{q},E^{q},\ldots $) and for their arguments $x,\xi ,t$. In the RHS of
(\ref{Radyushkin-ineq}), $q(x_{k})$ is the usual (forward) quark unpolarized
distribution taken at values
\begin{equation}
x_{1}=\frac{x+\xi }{1+\xi },\quad x_{2}=\frac{x-\xi }{1-\xi }\,.
\label{x-x1-x2-xi}
\end{equation}

In Ref.~\cite{DFJK-00} it was noticed that in the original derivation
of inequality (\ref{Radyushkin-ineq})  the contribution of GPD $E^{q}$
was ignored. The authors of Ref.~\cite{DFJK-00} have derived the correct
version of inequality (\ref{Radyushkin-ineq})
\begin{equation}
\left| H^{q}(x,\xi ,t)-\frac{\xi ^{2}}{1-\xi ^{2}}E^{q}(x,\xi ,t)\right|
\leq \sqrt{\frac{q(x_{1})q(x_{2})}{1-\xi ^{2}}}\,.  \label{ineq-DFJK-1}
\end{equation}

Later in Ref.~\cite{Pobylitsa-01} a stronger bound for $H^{q}$ and $E^{q}$
was obtained:
\[
\left[ H^{q}(x,\xi ,t)-\frac{\xi ^{2}}{1-\xi ^{2}}E^{q}(x,\xi ,t)\right]
^{2}
\]
\begin{equation}
+\left[ \frac{\sqrt{t_{0}-t}}{2m\sqrt{1-\xi ^{2}}}E^{q}(x,\xi ,t)\right]
^{2}\leq \frac{q(x_{1})q(x_{2})}{1-\xi ^{2}}  \label{ineq-P-01}\,.
\end{equation}
Here $m$ is the nucleon mass and $t_{0}$ is the maximal (negative) value of
the squared nucleon momentum transfer $t$
\begin{equation}
t_{0}=-\frac{4\xi ^{2}m^{2}}{1-\xi ^{2}},\quad t\leq t_{0}\leq 0\,.
\label{t-0-def}
\end{equation}

One has to keep in mind that the general positivity bound (\ref{start-ineq})
contains much more constraints on GPDs than the above examples
(\ref{ineq-DFJK-1}), (\ref{ineq-P-01}). The aim of this
paper is to extract from inequality (\ref{start-ineq}) as much information as possible.

In Ref.~\cite{Diehl-0101335} a classification of twist-2 GPDs was suggested
in terms of eight distributions
\begin{equation}
H,E,H_{T},E_{T},\tilde{H},\tilde{E},\tilde{H}_{T},\tilde{E}_{T}\,.
\label{GPDs-Diehl}
\end{equation}
Inequality (\ref{start-ineq}) defines some domain in the eight-dimensional
space of GPDs (\ref{GPDs-Diehl}). In this paper this ``allowed region'' is
studied.

The paper is organized as follows. In Section
\ref{helicity-amplitudes-section} the positivity condition is
formulated in terms of helicity amplitudes. In Section
\ref{solution-section} we find the region
in the eight-dimensional space of twist-two GPDs which is allowed
by the positivity constraints. In Section \ref{separate-GPDs-bound}
we show how to project this eight-dimensional ``allowed region'' onto
one axis corresponding to a single GPD, then we derive the general
inequality which can be applied to any GPD and to any linear combination
of GPDs. In Section
\ref{quark-GPDs-bounds} this general inequality is written in an explicit
form for various quark GPDs.
The main part of the paper is devoted to the quark GPDs.
In Section \ref{gluon-GPDs-bounds} similar inequalities for gluon GPDs are
presented.

\section{Positivity bound in terms of helicity amplitudes}
\label{helicity-amplitudes-section}

It is convenient to analyze the general positivity inequality
(\ref{start-ineq}) using ``helicity amplitudes'' (the term is not quite justified
but it is used for brevity) introduced in Ref.~\cite{Diehl-0101335}
\begin{equation}
A_{\lambda ^{\prime }\mu ^{\prime },\lambda \mu }=\left. \int \frac{dz^{-}}{
2\pi }e^{ixP^{+}z^{-}}\langle p^{\prime },\lambda ^{\prime }|O_{\mu ^{\prime
},\mu }(z)|p,\lambda \rangle \right| _{z^{+}=0,z^{\perp }=0}\,.
\label{A-amplitudes}
\end{equation}
Here $O_{\mu ^{\prime },\mu }$ are bilinear quark light-ray operators with
the polarization indices $\mu ,\mu ^{\prime }$. Next, $|p,\lambda \rangle $
is a nucleon state with momentum $p$ and polarization $\lambda $ (in the
sense of light-cone helicity states \cite{KS-70}), $P^{+}$ is the light-cone
component of vector $P=(p+p^{\prime })/2$. The explicit expressions for
$A_{\lambda ^{\prime }\mu ^{\prime },\lambda \mu }$
in terms of GPDs (\ref{GPDs-Diehl}) are listed in the Appendix.

With two values for each polarization index of $A_{\lambda ^{\prime }\mu
^{\prime },\lambda \mu }$ we have $2^{4}=16$ components but due to the
parity invariance \cite{Diehl-0101335}
\begin{equation}
A_{-\lambda ^{\prime },-\mu ^{\prime };-\lambda ,-\mu }=(-1)^{\lambda
^{\prime }-\mu ^{\prime }-\lambda +\mu }A_{\lambda ^{\prime }\mu ^{\prime
},\lambda \mu }  \label{helicity-reflection}
\end{equation}
only 8 components are independent. This is exactly the number of twist-2
GPDs (\ref{GPDs-Diehl}). There is a one-to-one linear correspondence between
the independent components of helicity amplitudes $A_{\lambda ^{\prime }\mu
^{\prime },\lambda \mu }$ and GPDs (\ref{GPDs-Diehl}). Therefore we
can study the positivity bound (\ref{start-ineq}) using
helicity amplitudes $A_{\lambda ^{\prime }\mu ^{\prime },\lambda \mu }$.
The result will be expressed in terms
of usual GPDs $H,E,\ldots $ (\ref{GPDs-Diehl}) at the final stage
of the work.
The
helicity amplitudes $A_{\lambda ^{\prime }\mu ^{\prime
},\lambda \mu }$ are useful
for the derivation of the positivity bounds
because the underlying inequality (\ref{start-ineq}) looks quite simple in
terms of $A_{\lambda ^{\prime }\mu ^{\prime },\lambda \mu }$:
\[
c_{M}^{(1)\ast }A_{MN}^{\dagger }(x,\xi ,t)c_{N}^{(2)}+c_{M}^{(2)\ast
}A_{MN}(x,\xi ,t)c_{N}^{(1)}
\]
\begin{equation}
+c_{M}^{(1)\ast }A_{MN}(x_{1},0,0)c_{N}^{(1)}+c_{M}^{(2)\ast
}A_{MN}(x_{2},0,0)c_{N}^{(2)}\geq 0\,.  \label{A-ineq}
\end{equation}
Here we combined the nucleon-quark polarization indices $\lambda ,\mu $ into
one multiindex $M=(\lambda ,\mu )$. The dagger in $A_{MN}^{\dagger }$ stands
for the Hermitean conjugation. As usual, the summation over repeated
indices is implied. Parameters $x_{1},x_{2}$ appearing in $A_{MN}(x_{1,2},0,0)$
are connected to arguments $x,\xi $ of $A_{MN}(x,\xi ,t)$ by relations
(\ref{x-x1-x2-xi}) and the region $x>|\xi |$ is implied.

In this paper inequality (\ref{A-ineq}) is analyzed in its general form with
arbitrary coefficients $c_{M}^{(k)}$. Allowing arbitrary coefficients
$c_{M}^{(k)}$ we extend the analysis to all kinds of GPDs of leading twist
including those with the quark helicity flip. It is interesting that
including the GPDs with the quark helicity flip in the analysis  we get new
constraints on helicity non-flip GPDs.

Inequality (\ref{A-ineq}) that is
valid for arbitrary coefficients
$c_{M}^{(k)}$ is nothing else but the positivity
condition for the following $8\times 8$ matrix:
\begin{equation}
\left(
\begin{array}{cc}
A(x_{1},0,0) & A^{\dagger }(x,\xi ,t) \\
A(x,\xi ,t) & A(x_{2},0,0)
\end{array}
\right) \geq 0\,.  \label{positvity-8-start}
\end{equation}
In the next section we shall transform this positivity condition
to a form more convenient for the practical work with GPDs.

\section{Solution of positivity constraint}
\label{solution-section}

The first step is to reduce the analysis of the positivity of the $8\times 8$
matrix (\ref{positvity-8-start}) to a problem involving only $4\times 4$
matrices. This can be done using the symmetry property
(\ref{helicity-reflection}). This symmetry allows us to make matrix $A_{MN}$ block
diagonal. Performing the transformation
\begin{equation}
c_{++}^{(k)}=\tilde{c}_{1}^{(k)}+\tilde{c}_{3}^{(k)}\,,\quad c_{--}^{(k)}=
\tilde{c}_{1}^{(k)}-\tilde{c}_{3}^{(k)}\,,
\end{equation}
\begin{equation}
c_{+-}^{(k)}=\tilde{c}_{2}^{(k)}+\tilde{c}_{4}^{(k)}\,,\quad c_{-+}^{(k)}=
\tilde{c}_{4}^{(k)}-\tilde{c}_{2}^{(k)}\,
\end{equation}
we obtain
\begin{equation}
c_{M}^{(k)\ast }A_{MN}c_{N}^{(l)}=\tilde{c}_{a}^{(k)\ast }\tilde{A}_{ab}
\tilde{c}_{b}^{(l)}\,.  \label{A-tilde-def}
\end{equation}
where the $4\times 4$ matrix $\tilde{A}_{ab}$ consists of two $2\times 2$
blocks $B_{1},B_{2}$
\begin{equation}
\tilde{A}=\left(
\begin{array}{cc}
B_{1} & 0 \\
0 & B_{2}
\end{array}
\right) \,.  \label{A-ab}
\end{equation}
The explicit expressions for matrix elements $A_{MN}$ and $\tilde{A}_{ab}$
are given in the Appendix. Since the $4\times 4$ matrix $\tilde{A}$ consists
of two $2\times 2$ blocks it has 8 nonzero components and this is exactly
the number of all twist-2 GPDs (\ref{GPDs-Diehl}).

In this representation the positivity condition (\ref{positvity-8-start})
results in two independent positivity constraints for $4\times 4$ matrices
\begin{equation}
C_{s}\equiv \left(
\begin{array}{cc}
B_{s}(x_{1},0,0) & B_{s}^{\dagger }(x,\xi ,t) \\
B_{s}(x,\xi ,t) & B_{s}(x_{2},0,0)
\end{array}
\right) \geq 0\quad (s=1,2)\,.  \label{C-s-positivity}
\end{equation}
The positivity of matrix $C_{s}$ automatically leads to the positivity of
its diagonal blocks

\begin{equation}
F_{s}(x_{k})=B_{s}(x_{k},0,0)\geq 0\,.
\end{equation}
Matrices $F_{s}$ are diagonal and correspond to the following combinations
of usual forward distributions
\begin{equation}
F_{s}=\left(
\begin{array}{cc}
q+\Delta _{L}q+2\varepsilon _{s}\Delta _{T}q & 0 \\
0 & q-\Delta _{L}q
\end{array}
\right)  \label{F1}
\end{equation}
\[
 \quad \quad (\varepsilon _{1}=-\varepsilon _{2}=1)\,.
\]
Here $q,\Delta _{L}q,\Delta _{T}q$ are, respectively, unpolarized,
longitudinally polarized, and transversity quark distributions. Thus we
reproduce the standard positivity bounds on parton distributions
\begin{equation}
\left| \Delta _{L}q\right| \leq q
\end{equation}
including the Soffer inequality \cite{Soffer-95}
\begin{equation}
2\left| \Delta _{T}q\right| \leq q+\Delta _{L}q\,.  \label{Soffer-ineq}
\end{equation}
Now we can replace (\ref{C-s-positivity}) by the following equivalent
condition:
\[
\left(
\begin{array}{cc}
F_{s}^{-1/2}(x_{1}) & 0 \\
0 & F_{s}^{-1/2}(x_{2})
\end{array}
\right) C_{s}\left(
\begin{array}{cc}
F_{s}^{-1/2}(x_{1}) & 0 \\
0 & F_{s}^{-1/2}(x_{2})
\end{array}
\right)
\]
\begin{equation}
=\left(
\begin{array}{cc}
1 & K_{s}^{+} \\
K_{s} & 1
\end{array}
\right) \geq 0  \label{K-positivity}
\end{equation}
where
\begin{equation}
K_{s}=F_{s}^{-1/2}(x_{2})B_{s}(x,\xi ,t)F_{s}^{-1/2}(x_{1})\,.
\label{K-explicit}
\end{equation}
It is straightforward to check that positivity condition
(\ref{K-positivity}) is equivalent to the following bound on the norm of matrix $K_{s}$
\begin{equation}
\left\| K_{s}\right\| \leq 1\,.
\label{K-norm-bound}
\end{equation}
Equivalently this can be reformulated as the positivity of the matrix
\begin{equation}
1-K_{s}^{\dagger }K_{s}\geq 0\,  \label{K-plus-K-bound}
\end{equation}
or in terms of matrix elements
\begin{equation}
\sum\limits_{j}|(K_{s})_{ij}|^{2}\leq 1\,,\quad
\sum\limits_{i}|(K_{s})_{ij}|^{2}\leq 1\,.
\end{equation}
Using the trace and the determinant of matrix $K_{s}^{\dagger }K_{s}$ we can
rewrite these constraints as follows
\begin{equation}
\mathrm{Tr}K_{s}^{\dagger }K_{s}\leq 2,\quad \mathrm{Tr}K_{s}^{\dagger
}K_{s}\leq 1+\det K_{s}^{\dagger }K_{s}\,.
\end{equation}
In our case of matrix $K_{s}$ (\ref{K-explicit}) these conditions take the
form
\begin{equation}
\mathrm{Tr}\left[ B_{s}^{\dagger }(x,\xi ,t)F_{s}^{-1}(x_{2})B_{s}(x,\xi
,t)F_{s}^{-1}(x_{1})\right] \leq 2\,,  \label{constriant-1}
\end{equation}
\[
\mathrm{Tr}\left[ B_{s}^{\dagger }(x,\xi ,t)F_{s}^{-1}(x_{2})B_{s}(x,\xi
,t)F_{s}^{-1}(x_{1})\right]
\]
\begin{equation}
 \leq 1+\frac{\left| \det B_{s}(x,\xi ,t)\right|
^{2}}{\det F_{s}(x_{1})\det F_{s}(x_{2})}\,.  \label{constriant-2}
\end{equation}
Next we note that the matrix $B_{s}(x,\xi ,t)$ is composed of the linear
combinations of various GPDs. The explicit expressions for the matrix
elements of $B_{s}$ in terms of GPDs (\ref{GPDs-Diehl})
 can be read from (\ref{A-ab}) using the
matrix elements $\tilde{A}_{ab}$ listed in the Appendix. Inserting these
expressions into inequalities (\ref{constriant-1}), (\ref{constriant-2}) one
can write explicit bounds on GPDs (\ref{GPDs-Diehl}). Note that the trace $\mathrm{Tr}\left[
B_{s}^{\dagger }F_{s}^{-1}(x_{2})B_{s}F_{s}^{-1}(x_{1})\right] $ is
quadratic in GPDs and the squared determinant $\left| \det B_{s}\right| ^{2}$
is quartic. Thus inequalities (\ref{constriant-1}), (\ref{constriant-2})
are polynomial inequalities for GPDs of order 2 and 4, respectively.
However,  instead of dealing with explicit expressions containing
polynomials of eight GPDs it is much more convenient to work
with the compact form of the positivity condition (\ref{K-norm-bound}).

\section{Bounds on separate GPDs and their linear combinations}
\label{separate-GPDs-bound}

With eight independent GPDs hidden in matrices $B_{s}(x,\xi ,t)$ the above
constraints (\ref{constriant-1}), (\ref{constriant-2}) are hardly tractable.
Therefore it is interesting to derive bounds for single GPDs. Since matrix
elements of $B_{s}(x,\xi ,t)$ are linear in GPDs we can represent any GPD
$G=H,E,\ldots $ (\ref{GPDs-Diehl}) in the form
\begin{equation}
G(x,\xi ,t)=\sum\limits_{s=1}^{2}\mathrm{Tr}\left[ L_{s}^{(G)}B_{s}(x,\xi ,t)
\right]  \label{G-L-B}
\end{equation}
where $L_{s}^{(G)}$ are $2\times 2$ matrices depending on the specific
choice of the GPD $G$. Hence the derivation of the bounds for single GPDs
reduces to the problem of the calculation of the maximum
\begin{equation}
\left| G(x,\xi ,t)\right| \leq \sum\limits_{s=1}^{2}\max_{B_{s}}\left|
\mathrm{Tr}\left[ L_{s}^{(G)}B_{s}(x,\xi ,t)\right] \right|
\end{equation}
under constraints (\ref{constriant-1}), (\ref{constriant-2}). Using relation
(\ref{K-explicit}) we can rewrite this as follows
\[
\max_{B_{s}}\left| \mathrm{Tr}\left[ L_{s}B_{s}(x,\xi ,t)\right] \right|
\]
\begin{equation}
=\max_{K_{s}:\left\| K_s\right\| \leq 1}\left| \mathrm{Tr}\left[
F_{s}^{1/2}(x_{1})L_{s}F_{s}^{1/2}(x_{2})K_{s}\right] \right|
\label{max-L-B}
\end{equation}
where the maximum in the RHS is taken with respect to matrices
$K_s$ obeying the constraint
$\left\| K_{s}\right\| \leq 1$. The general solution of this
problem for an arbitrary matrix $M$ is
\begin{equation}
\max_{K:\left\| K\right\| \leq 1}\left| \mathrm{Tr}(KM)\right| =\mathrm{Tr}
\left[ (M^{\dagger }M)^{1/2}\right] \,.
\label{max-K-M-general}
\end{equation}
Here $(M^{\dagger }M)^{1/2}$ should be understood as a function of the matrix.
Applying the general result (\ref{max-K-M-general})
to our case (\ref{max-L-B}) we find
\begin{equation}
\left| G(x,\xi ,t)\right| \leq \sum\limits_{s=1}^{2}\mathrm{Tr}\left( \left[
F_{s}(x_{1})L_{s}^{(G)}F_{s}(x_{2})L_{s}^{(G)\dagger }\right] ^{1/2}\right)
\,.  \label{G-bound-general}
\end{equation}
This inequality can be used for any GPD and any linear combination of GPDs.

\section{Bounds on quark GPDs}

\label{quark-GPDs-bounds}

\subsection{Inequalities for $H^{q}-\protect\xi ^{2}
{(1-\protect\xi^2)^{-1}}E^{q}$}

As an example let us consider a linear combination of GPDs $H^q$ and $E^q$ which
has a simple expression in terms of amplitudes $\tilde{A}_{ab}$ (\ref{A-ab}).
Using the equations of the Appendix we find
\begin{equation}
H^{q}-\frac{\xi ^{2}}{1-\xi ^{2}}E^{q}=\frac{1}{4\sqrt{1-\xi ^{2}}}\left(
\tilde{A}_{11}+\tilde{A}_{22}+\tilde{A}_{33}+\tilde{A}_{44}\right) \,.
\end{equation}
This structure corresponds to the following choice of matrices $L_{s}^{(G)}$
in Eq. (\ref{G-L-B})
\begin{equation}
L_{1}^{(G)}=L_{2}^{(G)}=\frac{1}{4\sqrt{1-\xi ^{2}}}\left(
\begin{array}{cc}
1 & 0 \\
0 & 1
\end{array}
\right) \,.
\end{equation}
With these matrices $L_{s}^{(G)}$ we obtain from
inequality (\ref{G-bound-general}):
\[
\left| H^{q}(x,\xi,t)-\frac{\xi ^{2}}{1-\xi ^{2}}E^{q}(x,\xi,t)\right|
\]
\begin{equation}
 \leq \frac{1}{4\sqrt{
1-\xi ^{2}}}\sum\limits_{s=1}^{2}\mathrm{Tr}\left( \left[
F_{s}(x_{1})F_{s}(x_{2})\right] ^{1/2}\right) \,.
\label{H-E-deriv}
\end{equation}
Below for brevity we omit the arguments $x,\xi,t$ of GPDs
and write the arguments $x_1,x_2$ of
the forward distributions as subscripts.

Using expression (\ref{F1}) for $F_{s}(x_{k})$ we can rewrite
inequality (\ref{H-E-deriv}) in the
form
\[
\left| H^{q}-\frac{\xi ^{2}}{1-\xi ^{2}}E^{q}\right| \leq \frac{1}{4\sqrt{
1-\xi ^{2}}}
\]
\[
\times
\left\{ \sqrt{\left( q+\Delta _{L}q-2\Delta _{T}q\right)
_{x_{1}}\left( q+\Delta _{L}q-2\Delta _{T}q\right) _{x_{2}}}\right.
\]
\[
+\sqrt{\left( q+\Delta _{L}q+2\Delta _{T}q\right) _{x_{1}}\left( q+\Delta
_{L}q+2\Delta _{T}q\right) _{x_{2}}}
\]
\begin{equation}
\left.
+2\sqrt{\left( q-\Delta
_{L}q\right) _{x_{1}}\left( q-\Delta _{L}q\right) _{x_{2}}}\,\right\} \,.
\label{H-E-strong}
\end{equation}
One can obtain a weaker bound by maximizing the RHS with respect to $\Delta
_{T}q$ in the range allowed by the Soffer inequality (\ref{Soffer-ineq})
\[
\left| H^{q}-\frac{\xi ^{2}}{1-\xi ^{2}}E^{q}\right|
\leq \frac1{2\sqrt{1-\xi ^{2}}}
\]
\[
\times\left\{
{\sqrt{\left(
q+\Delta _{L}q\right)_{x_{1}}\left( q+\Delta _{L}q\right) _{x_{2}}}}
\right.
\]
\begin{equation}
\left.
+\sqrt{
\left( q-\Delta _{L}q\right) _{x_{1}}\left( q-\Delta _{L}q\right) _{x_{2}} }
\,\right\}
\,.  \label{H-E-Delta-L}
\end{equation}

Another still weaker inequality can be derived by maximizing the RHS of
(\ref{H-E-Delta-L}) with respect to $\Delta _{L}q$
\begin{equation}
\left| H^{q}-\frac{\xi ^{2}}{1-\xi ^{2}}E^{q}\right| \leq \sqrt{\frac{
q(x_{1})q(x_{2})}{1-\xi ^{2}}}\,.  \label{H-E-weak}
\end{equation}
We have reproduced inequality (\ref{ineq-DFJK-1}).

Using this strategy one can obtain bounds for any GPD or any linear
combination of GPDs starting from the general inequality
(\ref{G-bound-general}). From the above example we see that there is a hierarchy
of inequalities:

1) strong inequalities where GPDs are bounded by combinations of all forward
quark distributions including the transversity distribution,

2) weaker inequalities without the forward transversity distribution,

3) still weaker inequalities where GPDs are bounded by only the unpolarized
forward distribution.

\subsection{Inequalities for $\tilde{H}^{q}
-{\protect\xi ^{2}}{(1-\protect\xi^2)^{-1}}\tilde{E}^{q}$}

Now let us turn to the combination of GPDs
\begin{equation}
\tilde{H}^{q}-\frac{\xi ^{2}}{1-\xi ^{2}}\tilde{E}^{q}=\frac{1}{4\sqrt{1-\xi
^{2}}}\left( \tilde{A}_{11}-\tilde{A}_{22}+\tilde{A}_{33}-\tilde{A}
_{44}\right) \,.
\end{equation}
This corresponds to the following matrices $L_{s}^{(G)}$ in Eq. (\ref{G-L-B})
\begin{equation}
L_{1}^{(G)}=L_{2}^{(G)}=\frac{1}{4\sqrt{1-\xi ^{2}}}\left(
\begin{array}{cc}
1 & 0  \\
0 & -1
\end{array}
\right) \,.
\end{equation}
Inserting these matrices $L_{1}^{(G)},L_{2}^{(G)}$ into the general
inequality (\ref{G-bound-general}) we arrive at the same expression in the
RHS as in the above section where the bounds for $H^{q}-\xi ^{2}(1-\xi
^{2})^{-1}E^{q}$ were derived. Therefore we have the same inequalities for
$H^{q}-\xi ^{2}(1-\xi ^{2})^{-1}E^{q}$ and $\tilde{H}^{q}-\xi ^{2}(1-\xi
^{2})^{-1}\tilde{E}^{q}$:
\[
\left| \tilde{H}^{q}-\frac{\xi ^{2}}{1-\xi ^{2}}\tilde{E}^{q}\right| \leq
\frac{1}{4\sqrt{1-\xi ^{2}}}
\]
\[
\times\left\{ \sqrt{\left( q+\Delta _{L}q-2\Delta
_{T}q\right) _{x_{1}}\left( q+\Delta _{L}q-2\Delta _{T}q\right) _{x_{2}}}
\right.
\]
\[
+\sqrt{\left( q+\Delta _{L}q+2\Delta _{T}q\right) _{x_{1}}\left( q+\Delta
_{L}q+2\Delta _{T}q\right) _{x_{2}}}
\]
\begin{equation}
\left. +2\sqrt{\left( q-\Delta
_{L}q\right) _{x_{1}}\left( q-\Delta _{L}q\right) _{x_{2}}}\,\right\} \,,
\end{equation}
\[
\left| \tilde{H}^{q}-\frac{\xi ^{2}}{1-\xi ^{2}}\tilde{E}^{q}\right| \leq
\frac1{2\sqrt{1-\xi ^{2}}}
\]
\[
\times\left\{
\sqrt{\left( q+\Delta _{L}q\right) _{x_{1}}\left( q+\Delta
_{L}q\right) _{x_{2}}}
\right.
\]
\begin{equation}
\left.
+\sqrt{\left( q-\Delta _{L}q\right) _{x_{1}}\left(
q-\Delta _{L}q\right) _{x_{2}}}\,
\right\}
\,,
\label{H-E-tilde-Delta-L}
\end{equation}
\begin{equation}
\left| \tilde{H}^{q}-\frac{\xi ^{2}}{1-\xi ^{2}}\tilde{E}^{q}\right| \leq
\sqrt{\frac{q(x_{1})q(x_{2})}{1-\xi ^{2}}}\,.
\end{equation}

\subsection{Inequalities for $E^{q}$}

In the derivation of inequalities for $E^{q}$ one deals with nondiagonal
matrices $L_{1}^{(G)},L_{2}^{(G)}$ in the RHS of (\ref{G-bound-general}).
Computing the trace in the RHS of (\ref{G-bound-general}) one obtains the
following bound
\[
|E^{q}|\leq \frac{m}{2\sqrt{t_{0}-t}}
\left\{ \sqrt{\left( q-\Delta _{L}q\right) _{x_{1}}\left( q+\Delta
_{L}q+2\Delta _{T}q\right) _{x_{2}}}
\right.
\]
\[
+\sqrt{\left( q+\Delta _{L}q+2\Delta
_{T}q\right) _{x_{1}}\left( q-\Delta _{L}q\right) _{x_{2}}}
\]
\[
+\sqrt{\left( q+\Delta _{L}q-2\Delta _{T}q\right) _{x_{1}}\left(
q-\Delta _{L}q\right) _{x_{2}}}
\]
\begin{equation}
\left.
+\sqrt{\left( q-\Delta _{L}q\right)
_{x_{1}}\left( q+\Delta _{L}q-2\Delta _{T}q\right) _{x_{2}}}\right\} .
\end{equation}
Maximizing the RHS with respect to $\Delta _{T}q$ we arrive at a weaker bound
\[
|E^{q}|\leq \frac{m}{\sqrt{t_{0}-t}}
\left[ \sqrt{\left( q+\Delta
_{L}q\right) _{x_{1}}\left( q-\Delta _{L}q\right) _{x_{2}}}
\right.
\]
\begin{equation}
\left.
+\sqrt{\left(
q-\Delta _{L}q\right) _{x_{1}}\left( q+\Delta _{L}q\right) _{x_{2}}}
\,\right]
\,.  \label{E-Delta-L-ineq}
\end{equation}
Next one can maximize the RHS of (\ref{E-Delta-L-ineq}) with respect $\Delta
_{L}q$
\begin{equation}
|E^{q}|\leq \frac{2m}{\sqrt{t_{0}-t}}\sqrt{q(x_{1})q(x_{2})}\,.
\label{E-weak-inequality}
\end{equation}
This coincides with
the bound for $E^{q}$ derived earlier in Ref.~\cite{DFJK-00} up to a typo
there \cite{Diehl-communication}. This bound also can be obtained directly
from inequality (\ref{ineq-P-01}).

\subsection{Inequalities for $\tilde{E}^{q}$}

Inequalities for $\tilde{E}^{q}$ differ from inequalities for $E^{q}$ only
by a factor of $\xi ^{-1}$ in the RHS:
\[
|\tilde{E}^{q}|\leq \frac{m}{2\xi \sqrt{t_{0}-t}}
 \left\{ \sqrt{\left( q-\Delta _{L}q\right) _{x_{1}}\left( q+\Delta
_{L}q+2\Delta _{T}q\right) _{x_{2}}}
\right.
\]
\[
+\sqrt{\left( q+\Delta _{L}q+2\Delta
_{T}q\right) _{x_{1}}\left( q-\Delta _{L}q\right) _{x_{2}}}
\]
\[
+\sqrt{\left( q+\Delta _{L}q-2\Delta _{T}q\right) _{x_{1}}\left(
q-\Delta _{L}q\right) _{x_{2}}}
\]
\begin{equation}
\left.
+\sqrt{\left( q-\Delta _{L}q\right)
_{x_{1}}\left( q+\Delta _{L}q-2\Delta _{T}q\right) _{x_{2}}}\,\right\} \,,
\end{equation}
\[
|\tilde{E}^{q}|\leq \frac{m}{\xi \sqrt{t_{0}-t}}\left[ \sqrt{\left( q+\Delta
_{L}q\right) _{x_{1}}\left( q-\Delta _{L}q\right) _{x_{2}}}
\right.
\]
\begin{equation}
\left.
+\sqrt{\left(
q-\Delta _{L}q\right) _{x_{1}}\left( q+\Delta _{L}q\right) _{x_{2}}}\,\right]
\,,  \label{E-tilde-Delta-L-ineq}
\end{equation}
\begin{equation}
|\tilde{E}^{q}|\leq \frac{2m}{\xi \sqrt{t_{0}-t}}\sqrt{q(x_{1})q(x_{2})}\,.
\label{E-tilde-weak-ineq}
\end{equation}

\subsection{Inequalities for $H^{q}$}

Using (\ref{G-bound-general}) one can derive the following bound for $H^{q}$
\[
|H^{q}|\leq \frac{1}{4}\left\{ \alpha ^{2}\left[ \sqrt{
f_{1}(x_{1})f_{1}(x_{2})}+\sqrt{f_{2}(x_{1})f_{2}(x_{2})}\right] ^{2}\right.
\]
\[
\left. +\beta ^{2}\left[ \sqrt{f_{1}(x_{1})f_{2}(x_{2})}+\sqrt{
f_{2}(x_{1})f_{1}(x_{2})}\right] ^{2}\right\} ^{1/2}
\]
\[
+\frac{1}{4}\left\{ \alpha ^{2}\left[ \sqrt{f_{3}(x_{1})f_{3}(x_{2})}+\sqrt{
f_{2}(x_{1})f_{2}(x_{2})}\right] ^{2}\right.
\]
\begin{equation}
\left. +\beta ^{2}\left[ \sqrt{f_{3}(x_{1})f_{2}(x_{2})}+\sqrt{
f_{2}(x_{1})f_{3}(x_{2})}\right] ^{2}\right\} ^{1/2}\,.
\label{H-bound-strong}
\end{equation}
Here we use a compact notation for the combinations of the usual (forward) parton
distributions
\[
f_{1}=q+\Delta _{L}q+2\Delta _{T}q\,,\quad f_{2}=q-\Delta _{L}q,
\]
\begin{equation}
\quad f_{3}=q+\Delta _{L}q-2\Delta _{T}q  \label{f-k-def}
\end{equation}
which appear in the diagonal matrices $F_{s}$ (\ref{F1}). Parameters $\alpha
,\beta $ are defined as follows
\begin{equation}
\alpha =\frac{1}{\sqrt{1-\xi ^{2}}},\quad \beta =\frac{\xi ^{2}}{1-\xi ^{2}}
\frac{2m}{\sqrt{t_{0}-t}}\,.  \label{alpha-beta-def}
\end{equation}
The weaker bound without transversity distribution $\Delta _{T}q$ looks as
follows
\[
|H^{q}|\leq \frac{1}{2}\left\{ \alpha ^{2}\left[ \sqrt{\left( q+\Delta
_{L}q\right) _{x_{1}}\left( q+\Delta _{L}q\right) _{x_{2}}}
\right.\right.
\]
\[
\left.\left.
+\sqrt{\left(
q-\Delta _{L}q\right) _{x_{1}}\left( q-\Delta _{L}q\right) _{x_{2}}}\,\right]
^{2}\right.
\]
\[
\left. +\beta ^{2}\left[ \sqrt{\left( q+\Delta _{L}q\right) _{x_{1}}\left(
q-\Delta _{L}q\right) _{x_{2}}}
\right.\right.
\]
\begin{equation}
\left.\left.
+\sqrt{\left( q-\Delta _{L}q\right)
_{x_{1}}\left( q+\Delta _{L}q\right) _{x_{2}}}\,\right] ^{2}\right\} ^{1/2}\,.
\label{H-bound-Delta-L}
\end{equation}
Maximizing the RHS with respect to $\Delta _{L}q$
we obtain
\begin{equation}
\left| H^{q}\right| \leq \sqrt{\left( 1+\frac{-t_{0}\xi ^{2}}{t_{0}-t}
\right) \frac{q(x_{1})q(x_{2})}{1-\xi ^{2}}}\,.  \label{H-bound-supernew}
\end{equation}
Note that inequality (\ref{H-bound-supernew}) is stronger than the bound for
$H^{q}$ derived in \cite{Pobylitsa-01} but still weaker than the obsolete
inequality (\ref{Radyushkin-ineq}).

\subsection{Inequalities for $\tilde{H}^{q}$}

Similarly we find for $\tilde{H}^{q}$

\[
|\tilde{H}^{q}|\leq \frac{1}{4}\left\{ \alpha ^{2}\left[ \sqrt{
f_{1}(x_{1})f_{1}(x_{2})}+\sqrt{f_{2}(x_{1})f_{2}(x_{2})}\right] ^{2}\right.
\]
\[
\left. +\left( \frac{\beta }{\xi }\right) ^{2}\left[ \sqrt{
f_{1}(x_{1})f_{2}(x_{2})}+\sqrt{f_{2}(x_{1})f_{1}(x_{2})}\right]
^{2}\right\} ^{1/2}
\]
\[
+\frac{1}{4}\left\{ \alpha ^{2}\left[ \sqrt{f_{3}(x_{1})f_{3}(x_{2})}+\sqrt{
f_{2}(x_{1})f_{2}(x_{2})}\right] ^{2}\right.
\]
\begin{equation}
\left. +\left( \frac{\beta }{\xi }\right) ^{2}\left[ \sqrt{
f_{3}(x_{1})f_{2}(x_{2})}+\sqrt{f_{2}(x_{1})f_{3}(x_{2})}\right]
^{2}\right\} ^{1/2}\,.  \label{H-tilde-bound-strong}
\end{equation}
Parameters $\alpha ,\beta $ are given in Eqs. (\ref{alpha-beta-def}) and
functions $f_{k}$ are the same as in Eqs. (\ref{f-k-def}). The weaker
inequality without the transversity distribution is
\[
|\tilde{H}^{q}|\leq \frac{1}{2}\left\{ \alpha ^{2}\left[ \sqrt{\left(
q+\Delta _{L}q\right) _{x_{1}}\left( q+\Delta _{L}q\right) _{x_{2}}}
\right.\right.
\]
\[
\left.\left.
+\sqrt{
\left( q-\Delta _{L}q\right) _{x_{1}}\left( q-\Delta _{L}q\right) _{x_{2}}}
\,\right] ^{2}\right.
\]
\[
\left. +\left( \frac{\beta }{\xi }\right) ^{2}\left[ \sqrt{\left( q+\Delta
_{L}q\right) _{x_{1}}\left( q-\Delta _{L}q\right) _{x_{2}}}
\right.\right.
\]
\begin{equation}
\left.\left.
+\sqrt{\left(
q-\Delta _{L}q\right) _{x_{1}}\left( q+\Delta _{L}q\right) _{x_{2}}}\,\right]
^{2}\right\} ^{1/2}
\end{equation}
and the bound without forward polarized distributions looks as follows
\begin{equation}
|\tilde{H}^{q}|\leq \sqrt{\frac{-t}{t_{0}-t}\frac{q(x_{1})q(x_{2})}{1-\xi
^{2}}}\,.  \label{H-tilde-ineq}
\end{equation}

\section{Inequalities for gluon GPDs}

\label{gluon-GPDs-bounds}

One can easily generalize the above inequalities for the case of gluon GPDs.
As noted in Ref.~\cite{Diehl-0101335} helicity amplitudes $A_{\lambda
^{\prime }\mu ^{\prime },\lambda \mu }$ without parton helicity flip (i.e.
with $\mu =\mu ^{\prime }$) are represented by the same expressions in terms
of GPDs in the quark and gluon cases. As a result the expressions for
$H,E,\tilde{H},\tilde{E}$ in terms of amplitudes $\tilde{A}_{ab}$ also have the
same form for quarks and gluons.

The difference between the quark and gluon cases comes from the following
sources:

1) in the gluon case the quark inequality (\ref{A-ineq}) should be replaced
by
\[
\frac{1}{\sqrt{1-\xi ^{2}}}\left[ c_{M}^{(1)\ast }A_{MN}^{\dagger }(x,\xi
,t)c_{N}^{(2)}+c_{M}^{(2)\ast }A_{MN}(x,\xi ,t)c_{N}^{(1)}\right]
\]
\begin{equation}
+c_{M}^{(1)\ast }A_{MN}(x_{1},0,0)c_{N}^{(1)}+c_{M}^{(2)\ast
}A_{MN}(x_{2},0,0)c_{N}^{(2)}\geq 0\,,
\end{equation}

2) the standard definition of the \emph{forward} gluon distribution contains
an extra factor of $x$ compared to the quark distributions,

3) we use the normalization conventions of Ref.~\cite{Diehl-0101335} for
gluon GPDs so that the forward limit of gluon GPDs differs from the standard
forward gluon unpolarized $g$ and polarized $\Delta _{L}g$ distributions
 by a factor of $x$:
\begin{equation}
H^{g}(x,0,0)=xg(x)\,,\quad \tilde{H}^{g}(x,0,0)=x\Delta _{L}g(x)\,.
\end{equation}

Combining all above factors we see that the transition from the quark case
to the gluon one reduces to the replacement
\[
\sqrt{q_{1}(x_{1})q_{2}(x_{2})}\rightarrow \sqrt{1-\xi ^{2}}\sqrt{x_{1}x_{2}}
\sqrt{g_{1}(x_{1})g_{2}(x_{2})}
\]
\begin{equation}
=\sqrt{x^{2}-\xi ^{2}}\sqrt{
g_{1}(x_{1})g_{2}(x_{2})}
\end{equation}
where $g_{k}=g,\Delta_Lg$ are gluon analogues of quark distributions
$q_{k}=q,\Delta_Lq$. Of course,
one has to keep in mind that the forward gluon transversity distribution
vanishes in the case of spin-1/2 hadrons which is considered here.

As a result we obtain
\[
\left| H^{g}-\frac{\xi ^{2}}{1-\xi ^{2}}E^{g}\right| \leq \frac{1}{2}\sqrt{
\frac{x^{2}-\xi ^{2}}{1-\xi ^{2}}}
\]
\[
\times \left[ \sqrt{\left( g+\Delta _{L}g\right) _{x_{1}}\left( g+\Delta
_{L}g\right) _{x_{2}}}
\right.
\]
\begin{equation}
\left.
+\sqrt{\left( g-\Delta _{L}g\right) _{x_{1}}\left(
g-\Delta _{L}g\right) _{x_{2}}}\,\right] \,,
\end{equation}

\begin{equation}
\left| H^{g}-\frac{\xi ^{2}}{1-\xi ^{2}}E^{g}\right| \leq \sqrt{\frac{
x^{2}-\xi ^{2}}{1-\xi ^{2}}g(x_{1})g(x_{2})}\,,  \label{H-E-g}
\end{equation}
\[
\left| \tilde{H}^{g}-\frac{\xi ^{2}}{1-\xi ^{2}}\tilde{E}^{g}\right| \leq
\frac{1}{2}\sqrt{\frac{x^{2}-\xi ^{2}}{1-\xi ^{2}}}
\]
\[
\times \left[ \sqrt{\left( g+\Delta _{L}g\right) _{x_{1}}\left( g+\Delta
_{L}g\right) _{x_{2}}}
\right.
\]
\begin{equation}
\left.
+\sqrt{\left( g-\Delta _{L}g\right) _{x_{1}}\left(
g-\Delta _{L}g\right) _{x_{2}}}\,\right] \,,
\end{equation}
\begin{equation}
\left| \tilde{H}^{g}-\frac{\xi ^{2}}{1-\xi ^{2}}\tilde{E}^{g}\right| \leq
\sqrt{\frac{x^{2}-\xi ^{2}}{1-\xi ^{2}}g(x_{1})g(x_{2})} \,,
\end{equation}
\[
|E^{g}|\leq \frac{m}{\sqrt{t_{0}-t}}\sqrt{x^{2}-\xi ^{2}}
\left[ \sqrt{\left(
g-\Delta _{L}g\right) _{x_{1}}\left( g+\Delta _{L}g\right) _{x_{2}}}
\right.
\]
\begin{equation}
\left.
+\sqrt{
\left( g+\Delta _{L}g\right) _{x_{1}}\left( g-\Delta _{L}g\right) _{x_{2}}}
\,\right] \,,
\end{equation}
\begin{equation}
|E^{g}|\leq \frac{2m\sqrt{x^{2}-\xi ^{2}}}{\sqrt{t_{0}-t}}\sqrt{
g(x_{1})g(x_{2})}\,,
\end{equation}
\[
|\tilde{E}^{g}|\leq \frac{m\sqrt{x^{2}-\xi ^{2}}}{\xi \sqrt{t_{0}-t}}\left[
\sqrt{\left( g-\Delta _{L}g\right) _{x_{1}}\left( g+\Delta _{L}g\right)
_{x_{2}}}
\right.
\]
\begin{equation}
\left.
+\sqrt{\left( g+\Delta _{L}g\right) _{x_{1}}\left( g-\Delta
_{L}g\right) _{x_{2}}}\,\right] \,,
\end{equation}
\begin{equation}
|\tilde{E}^{g}|\leq \frac{2m\sqrt{x^{2}-\xi ^{2}}}{\xi \sqrt{t_{0}-t}}\sqrt{
g(x_{1})g(x_{2})}\,,
\end{equation}
\[
|H^{g}|\leq \frac{\sqrt{x^{2}-\xi ^{2}}}{2}
 \left\{ \alpha ^{2}\left[ \sqrt{\left( g+\Delta _{L}g\right)
_{x_{1}}\left( g+\Delta _{L}g\right) _{x_{2}}}
\right.\right.
\]
\[
\left.\left.
+\sqrt{\left( g-\Delta
_{L}g\right) _{x_{1}}\left( g-\Delta _{L}g\right) _{x_{2}}}\,\right]
^{2}\right.
\]
\[
\left. +\beta ^{2}\left[ \sqrt{\left( g-\Delta _{L}g\right) _{x_{1}}\left(
g+\Delta _{L}g\right) _{x_{2}}}
\right.\right.
\]
\begin{equation}
\left.\left.
+\sqrt{\left( g+\Delta _{L}g\right)
_{x_{1}}\left( g-\Delta _{L}g\right) _{x_{2}}}\,\right] ^{2}\right\} ^{1/2}\,,
\end{equation}

\begin{equation}
\left| H^{g}\right| \leq \sqrt{x^{2}-\xi ^{2}}\sqrt{\left( 1
+\frac{-t_{0}\xi^{2}}{t_{0}-t}\right) \frac{g(x_{1})g(x_{2})}{1-\xi ^{2}}}\,,
\end{equation}
\[
|\tilde{H}^{g}|\leq \frac{\sqrt{x^{2}-\xi ^{2}}}{2}
 \left\{ \alpha ^{2}\left[ \sqrt{\left( g+\Delta _{L}g\right)
_{x_{1}}\left( g+\Delta _{L}g\right) _{x_{2}}}
\right.\right.
\]
\[
\left.\left.
+\sqrt{\left( g-\Delta
_{L}g\right) _{x_{1}}\left( g-\Delta _{L}g\right) _{x_{2}}}\,\right]
^{2}\right.
\]
\[
\left. +\left( \frac{\beta }{\xi }\right) ^{2}\left[ \sqrt{\left( g-\Delta
_{L}g\right) _{x_{1}}\left( g+\Delta _{L}g\right) _{x_{2}}}
\right.\right.
\]
\begin{equation}
\left.\left.
+\sqrt{\left(
g+\Delta _{L}g\right) _{x_{1}}\left( g-\Delta _{L}g\right) _{x_{2}}}\,\right]
^{2}\right\} ^{1/2}\,,
\end{equation}
\begin{equation}
|\tilde{H}^{g}|\leq \sqrt{x^{2}-\xi ^{2}}\sqrt{\frac{-t}{t_{0}-t}\frac{
g(x_{1})g(x_{2})}{1-\xi ^{2}}}\,\,.
\end{equation}
Parameters $\alpha ,\beta $ are given by Eqs. (\ref{alpha-beta-def}). Inequality
(\ref{H-E-g}) was derived earlier in Ref.~\cite{DFJK-00}.

\section{Conclusions}

In this paper we found the domain in the eight-dimensional space of twist-2
GPDs (\ref{GPDs-Diehl})
which is allowed by the positivity constraint (\ref{start-ineq}). This
region is described by polynomial inequalities (\ref{constriant-1}),
(\ref{constriant-2}). These inequalities mix various GPDs in a nontrivial way.
If one is interested in bounds on a single GPD then one has to project
this eight-dimensional domain onto the axis corresponding to the chosen
GPD. This projection leads to the general inequality (\ref{G-bound-general})
which can be applied to any GPD or any linear combination of GPDs.
The explicit inequalities for various quark GPDs are presented in
Section~\ref{quark-GPDs-bounds}
and for the gluon GPDs in Section~\ref{gluon-GPDs-bounds}.

The derivation of the positivity bounds presented here
ignored the problem of the renormalization. It is well known that in the case of forward parton
distributions the positivity can be violated at low normalization points
where parton distributions can essentially differ from the corresponding
structure functions associated with physical cross sections.
Similarly the positivity bounds derived for GPDs in this paper can be
also violated at low normalization points. In the case of forward
parton distributions the one-loop evolution to higher normalization points
is known to preserve all positivity properties. For
some special inequalities for GPDs
the stability with respect to the evolution was analyzed in Ref.~\cite{PST-99}. 
In Ref.~\cite{Pobylitsa-02} the one-loop evolution stability was demonstrated for a wider class
of positivity bounds including the inequalities considered in this paper.

I appreciate discussions with A.~Belitsky, J.C.~Collins, M.~Diehl,
L.~Frankfurt,
X.~Ji, M.~Kirch, N.~Kivel,
D.~M{\"u}ller, V.Yu.~Petrov, M.V.~Polyakov, A.V.~Radyushkin,
M.~Strikman and O.~Teryaev.

\appendix

\section{Helicity amplitudes}

Below we list the expressions for the amplitudes
$A_{\lambda ^{\prime }\mu ^{\prime },\lambda \mu }$ (\ref{A-amplitudes})
in terms of GPDs (\ref{GPDs-Diehl}). The derivation and the details can be found
in Ref.~\cite{Diehl-0101335}
\begin{equation}
A_{++,++}^{q}=\sqrt{1-\xi ^{2}}\left( \frac{H^{q}+\tilde{H}^{q}}{2}-\frac{
\xi ^{2}}{1-\xi ^{2}}\frac{E^{q}+\tilde{E}^{q}}{2}\right)
\,,
\end{equation}
\begin{equation}
A_{-+,-+}^{q}=\sqrt{1-\xi ^{2}}\left( \frac{H^{q}-\tilde{H}^{q}}{2}-\frac{
\xi ^{2}}{1-\xi ^{2}}\frac{E^{q}-\tilde{E}^{q}}{2}\right)
\,,
\end{equation}
\begin{equation}
A_{++,-+}^{q}=-\varepsilon \frac{\sqrt{t_{0}-t}}{2m}\frac{E^{q}-\xi \tilde{E}
^{q}}{2}
\,,
\end{equation}
\begin{equation}
A_{-+,++}^{q}=\varepsilon \frac{\sqrt{t_{0}-t}}{2m}\frac{E^{q}+\xi \tilde{E}
^{q}}{2}
\,,
\end{equation}
\begin{equation}
A_{++,+-}^{q}=\varepsilon \frac{\sqrt{t_{0}-t}}{2m}\left( \tilde{H}
_{T}^{q}+(1-\xi )\frac{E_{T}^{q}+\tilde{E}_{T}^{q}}{2}\right)
\,,
\end{equation}
\begin{equation}
A_{-+,--}^{q}=\varepsilon \frac{\sqrt{t_{0}-t}}{2m}\left( \tilde{H}
_{T}^{q}+(1+\xi )\frac{E_{T}^{q}-\tilde{E}_{T}^{q}}{2}\right)
\,,
\end{equation}
\[
A_{++,--}^{q}=\sqrt{1-\xi ^{2}}\left( H_{T}^{q}+\frac{t_{0}-t}{4m^{2}}\tilde{
H}_{T}^{q}
\right.
\]
\begin{equation}
\left.
-\frac{\xi ^{2}}{1-\xi ^{2}}E_{T}^{q}+\frac{\xi }{1-\xi ^{2}}
\tilde{E}_{T}^{q}\right)
\,,
\end{equation}
\begin{equation}
A_{-+,+-}^{q}=-\sqrt{1-\xi ^{2}}\frac{t_{0}-t}{4m^{2}}\tilde{H}_{T}^{q}
\,.
\end{equation}
Here $\varepsilon =\pm 1$ is a sign factor whose value is not important for
the derivation of the bounds on GPDs and $t_{0}$ is defined in
(\ref{t-0-def}).

Using the above expressions for $A_{\lambda ^{\prime }\mu ^{\prime },\lambda
\mu }$ one finds the following nonzero matrix elements $\tilde{A}
_{ab}$ (\ref{A-tilde-def})
\[
\tilde{A}_{11}=\sqrt{1-\xi ^{2}}\left[ H^{q}+\tilde{H}^{q}+2H_{T}^{q}
\right.
\]
\begin{equation}
\left.
+\frac{
-\xi ^{2}\left( E^{q}+\tilde{E}^{q}+2E_{T}^{q}\right) +2\xi \tilde{E}_{T}^{q}
}{1-\xi ^{2}}+\frac{t_{0}-t}{2m^{2}}\tilde{H}_{T}^{q}\right]
\,,
\end{equation}
\[
\tilde{A}_{22}=\sqrt{1-\xi ^{2}}\left[ H^{q}-\tilde{H}^{q}
\right.
\]
\begin{equation}
\left.
-\frac{\xi ^{2}}{
1-\xi ^{2}}\left( E^{q}-\tilde{E}^{q}\right) +\frac{t_{0}-t}{2m^{2}}\tilde{H}
_{T}^{q}\right]
\,,
\end{equation}
\begin{equation}
\tilde{A}_{12}=\varepsilon \frac{\sqrt{t_{0}-t}}{m}\left[ \tilde{H}
_{T}^{q}+(1-\xi )\frac{E_{T}^{q}+\tilde{E}_{T}^{q}}{2}+\frac{E^{q}-\xi
\tilde{E}^{q}}{2}\right]
\,,
\end{equation}
\[
\tilde{A}_{21}=\varepsilon \frac{\sqrt{t_{0}-t}}{m}\left[ -\tilde{H}
_{T}^{q}-(1+\xi )\frac{E_{T}^{q}-\tilde{E}_{T}^{q}}{2}
\right.
\]
\begin{equation}
\left.
-\frac{E^{q}+\xi
\tilde{E}^{q}}{2}\right]
\,,
\end{equation}
\[
\tilde{A}_{33}=\sqrt{1-\xi ^{2}}\left[ H^{q}+\tilde{H}^{q}-2H_{T}^{q}
\right.
\]
\begin{equation}
\left.
+\frac{
\xi ^{2}(-E^{q}-\tilde{E}^{q}+2E_{T}^{q})-2\xi \tilde{E}_{T}^{q}}{1-\xi ^{2}
}
-\frac{t_{0}-t}{2m^{2}}\tilde{H}_{T}^{q}\right]
\,,
\end{equation}
\[
\tilde{A}_{44}=\sqrt{1-\xi ^{2}}\left[ H^{q}-\tilde{H}^{q}-\frac{\xi ^{2}}{
1-\xi ^{2}}\left( E^{q}-\tilde{E}^{q}\right)
\right.
\]
\begin{equation}
\left.
-\frac{t_{0}-t}{2m^{2}}\tilde{H}
_{T}^{q}\right]
\,,
\end{equation}
\begin{equation}
\tilde{A}_{34}=\varepsilon \frac{\sqrt{t_{0}-t}}{m}\left[ \tilde{H}
_{T}^{q}+(1-\xi )\frac{E_{T}^{q}+\tilde{E}_{T}^{q}}{2}-\frac{E^{q}-\xi
\tilde{E}^{q}}{2}\right]
\,,
\end{equation}
\[
\tilde{A}_{43}=-\varepsilon \frac{\sqrt{t_{0}-t}}{m}\left[ \tilde{H}
_{T}^{q}+(1+\xi )\frac{E_{T}^{q}-\tilde{E}_{T}^{q}}{2}
\right.
\]
\begin{equation}
\left.
-\frac{E^{q}+\xi
\tilde{E}^{q}}{2}\right]
\,.
\end{equation}

\end{document}